# Lithospheric loading model for large impact basin where mantle plug presents

Qingyun Deng[1], Zhen Zhong[2], Mao Ye[1*], Wensong Zhang[1], Denggao Qiu[1], Chong Zheng[1], Jianguo Yan[1], Fei Li[1,3], Jean-Pierre Barriot[1,4]

[1]State Key Laboratory of Information Engineering in Surveying, Mapping and Remote Sensing, Wuhan University, Wuhan, China,

[2]School of Physics and Electronic Science, Guizhou Normal University, Guiyang, China,

[3]Chinese Antarctic Center of Surveying and Mapping, Wuhan University, Wuhan, China,

[4]Geodesy Observatory of Tahiti, Tahiti, French Polynesia

*Corresponding author: Mao Ye (mye@whu.edu.cn)

**Key points**

1. Post-impact crustal structure of large impact basin inferred from crustal thickness model is incorporated with lithospheric loading model.

2. Compared with density variation, mantle loading is the key to constructing lithosphere model at the impact basin.

3. Elastic thickness at Argyre and Isidis basins is estimated as 47.3 and 74.2 km, respectively.

**Abstract**

Lithosphere is an outer rigid part of the terrestrial body, usually consisting of the crust and part of the mantle. Characterizing the physical properties of the lithosphere is critical in investigating its evolution. By modeling mass-related loads within the lithosphere, physical parameters such as the elastic thickness of the lithosphere can be inferred from gravity and topography data. In the impact basin region, however, the low topography-gravity correlation and the sharp change in admittance from negative to positive within a narrow spheric harmonic degree make this model inapplicable. In this work, we incorporated mantle uplift structures commonly formed in impact basin regions into the lithospheric loading model. The crustal-mantle boundary of this mantle uplift structure is inferred from the global crustal thickness model. The gravity anomaly of the deflected lithosphere is calculated at the surface and crustal-mantle boundary, then the theoretical gravity admittance and correlation can be compared with the observed data. We sampled parameters using this mantle loading model at Argyre and Isidis basin on Mars with a novel crustal thickness model from the InSight mission. Our work suggests that proper modelling of the impact-induced load is critical to understanding the physical properties of the planetary lithosphere in the basin region.

**Plain Language Summary**

The lithosphere is defined as the outer rigid part of a planet. In the investigation of lithospheric physical properties, we usually treat the lithosphere as a thin elastic shell and calculate its deformation. Such deformation is measured by topography and gravity data. Theoretical transfer function and correlation between gravity and topography can be compared with observed data, thus deducing the physical properties of the lithosphere. However, this approach is difficult to implement in the impact basin region due to complex loading situations following impact processes, such as mass concentration (mascon) beneath the basin center. In this work, we modeled mascon features under impact basins as a mantle plug structure. We considered the mantle uplift as part of the net loading on the lithosphere, and inferred its structure from a global crustal thickness model. This mantle loading model is examined in large impact basins on Mars with different crustal thickness models.

## 1. Introduction

Thin elastic shell lithospheric model is widely used in geophysical studies of the Earth and other terrestrial bodies (Turcotte et al., 1981). This model assumes that the planetary lithosphere acts as an elastic shell that deflects when mass-related loads are imposed. The magnitude of deflection depends on the applied loads and physical properties of the lithosphere (such as elastic thickness). If this deflection changes the relief at a density interface, it will produce gravity anomaly. Therefore, deflection measurements can be made using gravity and topography data. Elastic thickness can be used to characterize lithospheric physical properties (Watts, 2001), which are mainly influenced by temperature structure and therefore closely related to the evolution of the entire planet.

The interpretation of gravity and topography data is usually performed in the spherical harmonic (SH) domain. Degree-dependent correlation and admittance (transfer function) between gravity and topography contain information about loads in lithosphere. Various types of loading models have been proposed. The simplest loading model is that only topographic loading (surface loading) is considered (e.g., Turcotte et al., 1981). The mixed loading model further takes into account the effect of sub-surface loading (e.g., Forsyth, 1985). Assuming that the surface and sub-surface loading is perfect in phase is a good approximation for regions where the correlation between gravity and topography is high. Nevertheless, geological processes such as erosion may cause surface and sub-surface loads out of phase (Forsyth, 1985). In this case, an additional phase factor should be involved in calculating the theoretical admittance function (Ding et al., 2019). McGovern et al. (2002) used a mass sheet approximation for the sub-surface loading, which differs from the density interface approximation of sub-surface loading suggested by Forsyth (1985). This model is further improved by adopting finite amplitude correction instead of mass sheet approximation. Belleguic et al. (2005) proposed a numerical rigorous method for calculating the loads induced by lateral density variation in the lithosphere and relief variation at the density interface. This research also ascribed geological significance to load ratio (f) by considering these possible loading types in the Martian lithosphere: magmatic intrusions within the crust or negative density perturbation in the mantle. Zhong et al. (2022) developed a spherical harmonic method of Belleguic et al. (2005), but with opposite signs in the definition of load ratio.

One issue that is still worth noting is proper modeling the load at impact basins. Presence of basin-related mascon (mass concentration) features usually results in a low or even negative correlation, and the admittance spectra show a pattern of transition from large negative values at low degrees to positive values at high degrees. This transition pattern may be related to large-scale, depressed topography, and positive gravity anomaly mascon structures. As one of the causes of basin-related mascon, the in-filling process after the end of loading (when the lithosphere is stronger) could also reduce the correlation between gravity anomaly and topography (Andrews-Hanna, 2013). Moreover, mantle plug structures always form during the impact process, elevating the crustal-mantle boundary (CrMB) and contributing to local gravity anomaly (Melosh et al., 2013, Freed et al., 2014, Johnson et al., 2016).

In previous studies considering infilling or/and mantle loading in the impact basin region, post-impact and pre-infilling crustal structures were almost inferred from topographic data, such as using basin scale relationships (Zhong et al., 2018, 2019; Ding et al., 2019), impact simulation results (Mancinelli et al., 2015), or initial isostatic hypothesis (Searls et al., 2006; Ritzer and Hauck, 2009). Inferring the sub-surface mantle structure from the isostatic hypothesis requires information on its original basin shape and degree of isostatic compensation (e.g., Ritzer and Hauck, 2009), but neither is well constrained for the post-impact state. Moreover, the impact process drastically changes the crustal structure, and this post-impact crustal structure is substantially influenced by temperature gradient, crustal thickness, and other factors (Freed et al., 2014). Therefore, the magnitude and shape of the mantle uplift and topographic depression soon after the impact are difficult to correlate. By modeling the mantle plug independently of topography, such as using a global crustal thickness model, it may be possible to provide a more realistic post-impact crustal structure. The uplifted CrMB with respect to the sub-surface equipotential interface leads to a high-density mantle replacing the low-density crust, which will exert pressure on the lithosphere. This suggests that uplifted-mantle plug structures should be incorporated with net vertical pressure into the deflection equation (see Section 2).

n this work, we propose a lithospheric loading model that is integrated with the post-impact mantle plug inferred from a global crustal thickness model. In order to integrate this post-impact mantle plug loading into the calculation of net vertical pressure (q), CrMB relief of the post-impact mantle plug is inferred from crustal thickness models, which avoids the assumption of the original basin shape and isostatic compensation. The deflection of the lithosphere (w) is calculated from net vertical pressure (q), changing initial topography (hi) and pre-deflection CrMB (mi) relief to form present-day topography (h) and post-deflection CrMB relief (m). The finite amplitude method (Wieczorek and Phillips, 1998) is used to calculate the gravity anomaly generated by topography (h) with respect to surface geoid and post-deflection CrMB relief (m) with respect to sub-surface equipotential interface. Calculation is carried out in the Argyre and Isidis basins on Mars. The implications of this newly proposed loading model in estimating elastic thickness are discussed.

**2. Model description**

Considering the lithosphere as a thin elastic shell, it will be flexural deformed under applied loads. The constitutive equation of the flexural deflection of the thin elastic shell is given by (Kraus, 1967; Turcotte et al., 1981):

$$D\nabla^6 w + 4D\nabla^4 w + ET_e R^2 \nabla^2 w + 2ET_e R^2 w = R^4(\nabla^2 + 1 - v)q \tag{1}$$

in which $D$ is flexural rigidity:

$$D = \frac{ET_e^3}{12(1-v^2)} \tag{2}$$

and $w$ denotes vertical deflection of the elastic lithosphere (positive downward). The flexural rigidity ($D$) is calculated by elastic modulus ($E$), Poisson ratio ($v$), and thickness of the shell (lithospheric elastic thickness, $T_e$). $R$ is the mean value of the planetary radius. $q$ is the vertical net pressure (measured positive downward). $\nabla^2$ is the Laplacian operator.

Load is the pressure exerted on the lithosphere. Mass from topography adds positive pressure to the lithosphere, causing the lithosphere to deflect. When lithosphere deflection perturbs the subsurface density contrast interface, such as the crustal-mantle boundary, changes in density interface relief can counterbalance part of the initial positive pressure. The net vertical pressure is expressed as:

$$q = g_1 \rho_c h - g_2(\rho_m - \rho_c)w \tag{3}$$

where $g_1$ and $g_2$ are gravitational acceleration at the surface geoid and sub-surface equipotential interface. $\rho_c$ and $\rho_m$ are crustal density and mantle density, respectively. $h$ is the present-day topography height with respect to the geoid (measured positively upward). Equation 3 is the net vertical pressure considering only *topography loading* (Figure 1a). To obtain deflection ($w$) as a function of topography ($h$), these variables ($w$ and $h$) are expanded into spherical harmonics. The ratio of $w$ to $h$ at each SH degree depends on the physical properties of the lithosphere, and can be inferred from gravity-topography admittance function. Therefore, by comparing observed and theoretical admittance, physical properties of the lithosphere, such as elastic thickness, can be inferred.

In most of the loading models previously proposed, the change in CrMB relief is assumed to be caused solely by the flexural deflection of the lithosphere (Figure 1a). Nevertheless, mantle plug structures formed during the impact process also increase CrMB relief relative to the initial level of equipotential (Melosh et al., 2013). Changes in lateral density within the lithosphere also affect net vertical pressure. At this stage, the net vertical pressure applied to the lithosphere includes depressed basal topography, uplifted CrMB relief, and density variation. Figure 1b describes the lithosphere state before and after deflection in such a case. We define depressed basal topography after impact as $h_i$, which may also be referred to as initial topography loading. The elevated CrMB relief with respect to the sub-surface equipotential is introduced as $m_i$ (also measured positively upwards) to reflect the post-impact mantle plug structures. The density variation is modeled with a density anomaly layer with a thickness of $B$ and density anomaly of $\Delta\rho$. $h_i$ and $m_i$ represent topography and CrMB relief before deflection occurs, while $h$ and $m$ represent present-day topography and post-deflection CrMB relief.

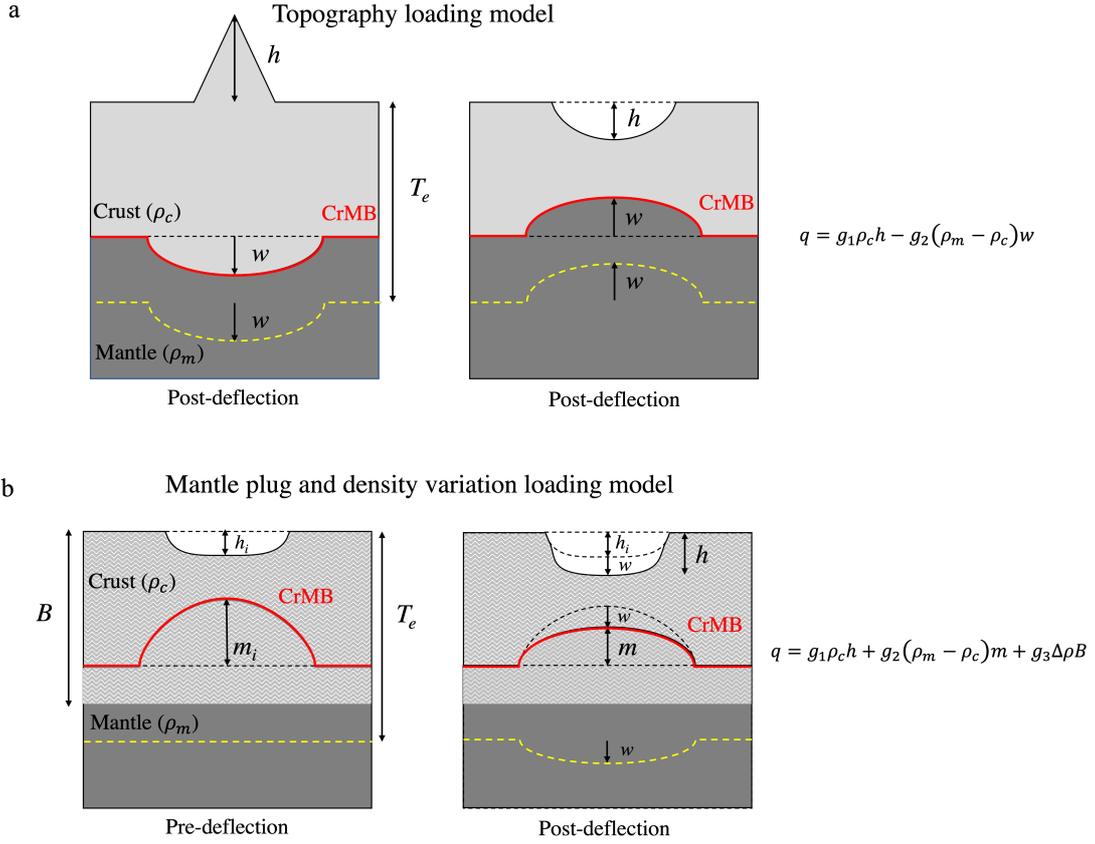

**Figure 1.** Schematic diagram of loading model. $h$ is present-day observed topography and $T_e$ is the elastic thickness of the lithosphere. (a) Two *topography loading* models ($h > 0$ and $h < 0$) showing the lithosphere after deflection. Changes in CrMB relief (red solid line) are assumed to be caused solely by the flexural deflection of the lithosphere ($w$, yellow dash line). The net vertical pressure is expressed as $q = g_1\rho_c h - g_2(\rho_m - \rho_c)w$. The left panel show the case of $h > 0$ and $w > 0$, while the right panel shows the case of $h < 0$ and $w < 0$. (b) Pre- and post-deflection of *mantle loading model with density variation*. $h_i$ is initial topography loading while $m_i$ is the post-impact pre-deflection CrMB relief. The thickness of the density variation layer is $B$. Due to deflection, CrMB relief changes from $m_i$ (pre-deflection after impact) to $m$ (post-deflection). Post-deflection CrMB relief ($m$) takes into account both $m_i$ and $w$. The net vertical pressure considers loading from topography, mantle plug, and density variation: $q = g_1\rho_c h + g_2(\rho_m - \rho_c)m + g_3\Delta\rho B$

The relationship between $m$, $m_i$, and $w$ is given by:

$$m = m_i - w \tag{4}$$

For $h$, $h_i$, and $w$, we have:

$$h = h_i - w \tag{5}$$

At this time (after deflection occur), the net vertical pressure for the *mantle loading model with density variation* is expressed as:

$$q = g_1\rho_c h + g_2(\rho_m - \rho_c)m + g_3\Delta\rho B \tag{6}$$

in which $g_3$ is gravity acceleration at the base of *B*. We assume that the density of the uplift mantle equals the surrounding mantle. Net vertical pressure from post-deflection topography, mantle plug, and density variation loading are expressed as $g_1\rho_c h$, $g_2(\rho_m - \rho_c)m$, and $g_3\Delta\rho B$ respectively.

Ritzer and Hauck (2009) used a similar approach to model mantle plug load. The pre-loading and after-loading crustal thickness anomaly (δc) in Ritzer and Hauck (2009) is similar to the pre- and post-deflection CrMB relief ($m_i$ and *m*) in our work. In Ritzer and Hauck (2009), crustal thickness anomaly calculations require assumptions about the original basin shape and compensation ratio.

Following Belleguic et al. (2005) and Zhong et al. (2022), a load parameter that indicates the relation between initial topography loading and density variation loading is introduced as:

$$f = -\frac{\Delta\rho B}{\rho_c h_i} \tag{7}$$

Substituting Equation 4, 5, 6, and 7 to Equation 1, we have the density variation (Δρ) and post-deflection CrMB relief (*m*) as (in spherical harmonics):

$$\Delta\rho = -\frac{\rho_c f}{B}\left(\frac{1+\alpha g_2(\rho_m-\rho_c)+\alpha g_1\rho_c}{1+\alpha g_2(\rho_m-\rho_c)+f\alpha g_3\rho_c}h + \frac{\alpha g_2(\rho_m-\rho_c)}{1+\alpha g_2(\rho_m-\rho_c)+f\alpha g_3\rho_c}m_i\right) \tag{8}$$

$$m = \frac{(f\alpha g_3\rho_c - \alpha g_1\rho_c)}{1+\alpha g_2(\rho_m-\rho_c)+f\alpha g_3\rho_c}h + \frac{(1+f\alpha g_3\rho_c)}{1+\alpha g_2(\rho_m-\rho_c)+f\alpha g_3\rho_c}m_i \tag{9}$$

The derivation of Equation 8 and Equation 9 is referred in Text S1. The compensation ratio $\alpha$ for each SH degree *n* is expressed as:

$$\alpha = \frac{n(n+1)-1+v}{\sigma[n^3(n+1)^3 - 4n^2(n+1)^2] + \tau[n(n+1)-2]} \tag{10}$$

and

$$\sigma = \frac{D}{R^4} = \frac{ET_e^3}{12R^4(1-v^2)} \tag{11}$$

$$\tau = \frac{ET_e}{R^2} \tag{12}$$

The post-deflection CrMB relief ($m$) is solved by $m_i$ and $w$ (Equation 9) and the gravity anomaly is calculated with density contrast of $\rho_m - \rho_c$. Gravity anomaly from surface topography is calculated with topographic height $h$ (reference to surface geoid) with $\rho_c + \Delta\rho$. These two gravity anomalies are added together to compute the theoretical gravity anomaly and gravity-topography admittance/correlation. Lithospheric parameters can be inferred by comparing between theoretical and observed admittance.

Since the post-impact pre-deflection CrMB relief of the mantle plug structure ($m_i$) is not directly observable, we used the crustal thickness model and topography model to infer this pre-deflection CrMB relief ($m_i$). Although the crustal thickness model is also derived using gravity and topography data, our method avoids the assumption of the original basin shape and compensation state in deriving pre-deflection CrMB relief ($m_i$).

### 3. Applications to impact basins on Mars

Next, we apply the mantle loading model to four large impact basins on Mars (Figure 2). They are Hellas (70°E, 42.7°S, D~1940 km), Argyre (316°E, 49.7°S, D~1700 km), Isidis (87°E, 12°N, D~1500 km), and Utopia (114°E, 42°N, D~2200 km). The Argyre and Hellas basins are located in the southern highlands, while Utopia is in the northern lowlands. The Isidis basin is located at the junction of the north-south dichotomy.

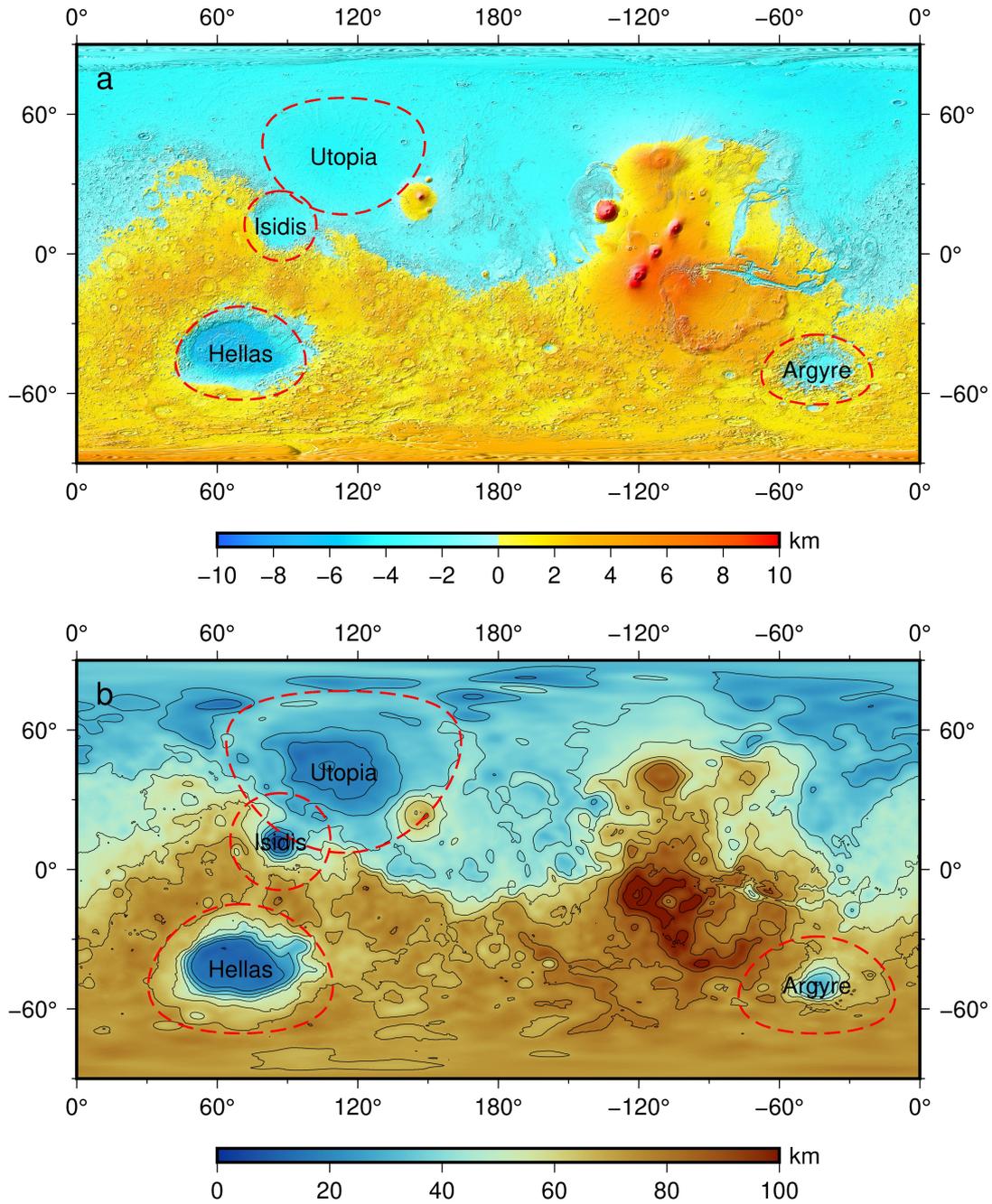

**Figure 2.** (a) Topography of Mars derived from the 2600-degree spherical harmonic shape model, referenced to geoid ($a$=3395.428 km, $b$=3377.678 km, reference potential $W_0$=12654875 m$^2$/s$^2$, Ardalan et al., 2010). (b) Crustal thickness model of Mars (Wieczorek, 2022) with $\rho_{c,south}$=2600 kg/m$^3$. Four large impact basins are indicated by red dashed circles.

3.1 Data

We used the 120-degree gravity field model of Mars jgmro_120f_sha model (Konopliv et al., 2020) and the MarsTopo719 shape model (Wieczorek, 2007) to calculate the observed admittance.

Topography model up to 2600 degrees (Wieczorek, 2007) is presented in Figure 2 for illustration purposes. The degree-dependent admittance function $z(l)$ and correlation $\gamma(l)$ between gravity and topography are calculated from their power and cross-power spectra (Wieczorek and Simons, 2005):

$$z(l) = \frac{S_{gh}(l)}{S_{hh}(l)} \tag{13}$$

$$\gamma(l) = \frac{S_{gh}(l)}{\sqrt{S_{gg}(l)S_{hh}(l)}} \tag{14}$$

in which $S_{hh}(l)$ and $S_{gg}(l)$ are the power spectra of topography and gravity, respectively. $S_{gh}(l)$ is the cross-power spectra between gravity and topography. The admittance error of $\sigma_z(l)$ is estimated as:

$$\sigma_z(l) = \left|\frac{z(l)}{\gamma(l)}\right| \sqrt{\frac{1-\gamma(l)^2}{2l}} \tag{15}$$

The localization window function is designed to include regions covering the entire basin and to ensure >99% concentration factor for the localized taper (Wieczorek and Simons, 2005). Depending on their size, we set the angular radius of the localization window for the Hellas, Argyre, Isidis, and Utopia basins as 20, 15, 15, and 25, respectively. If we desired >99% concentration factor for the taper, these localization window widths $l_{win}$ are obtained as 13, 17, 17, and 10, respectively.

Modeled (theoretical) admittance is obtained by summing gravity anomalies due to present-day topography (*h*) and post-deflection CrMB relief (*m*), and calculating together with the topography model. In fitting observed and modeled admittance, the same localization window is applied to both observed and modeled admittance.

The localized admittance for these four basins is shown in Figure 3. The error bars show three times the error for conservative estimation of lithospheric parameters in our analysis. We found that the Argyre (Figure 3a) and Isidis (Figure 3b) show the pattern of typical mascon characteristic: a sharp change from large negative to large positive admittance values within a narrow SH range. In the Hellas (Figure 3c) and Utopia (Figure 3d) basins, this mascon feature is not significant and distinguishable. The inversion results of our work (Section 4) show that the applicability of the mantle loading model is related to the prominence of the mascon admittance feature in the observed gravity admittance. Argyre and Isidis will be used as examples in the main text.

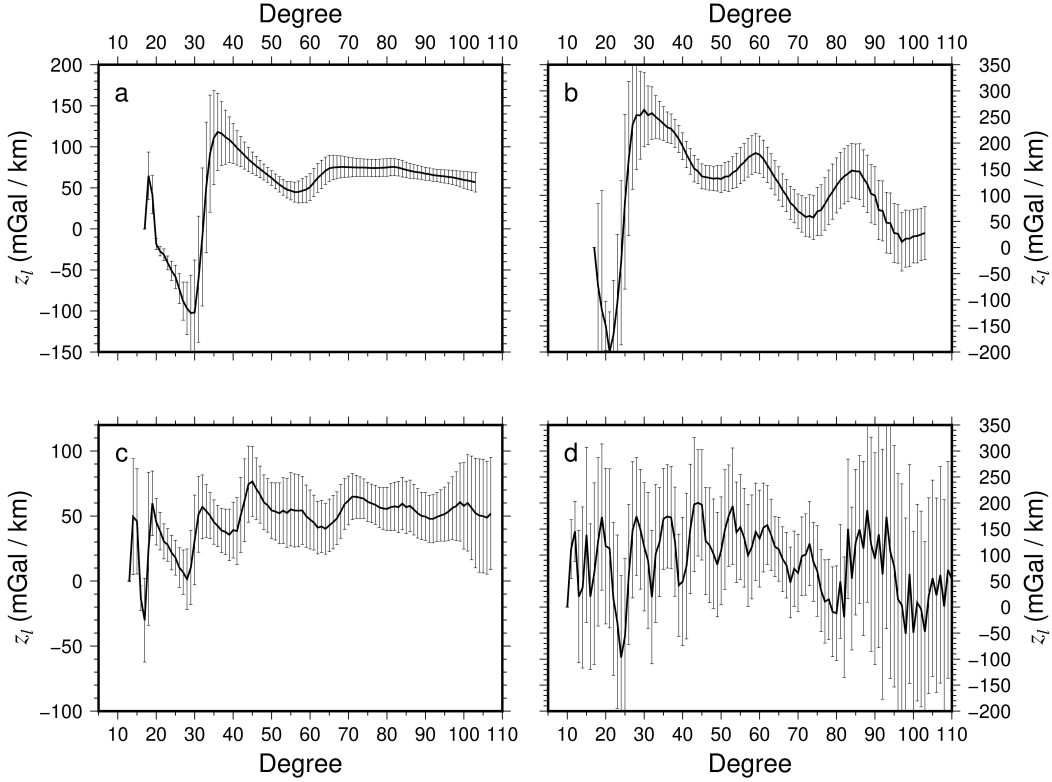

**Figure 3**. Localized admittance of (a) Argyre, (b) Isidis, (c) Hellas, and (d) Utopoa basin from gravity field model (Konopliv et al., 2020) and topography model (Wieczorek, 2007), referred as observed admittance. The error bars show three times the admittance error from Equation 15.

Crustal thickness models from the InSight mission (Wieczorek et al., 2022, here after the W2022 model) are used to infer pre-deflection CrMB relief. This series of crustal thickness models reveals a hemispheric dichotomy in crustal thickness. Average crustal thickness is 37 km for the northern lowlands and 63 km for the southern highlands. In deriving the global crustal thickness model, the crustal density of Martian northern lowlands is assumed to be 2900 kg/m³, and four crustal density values for southern highlands ($\rho_{c,south}$) are assigned as 2600, 2700, 2800, and 2900 kg/m³, respectively. Mantle density is set at 3382 kg/m³.

3.2 Forward calculation

In this section, we perform forward calculation of the proposed mantle loading model to examine parameter sensitivity. To study the sensitivity of elastic thickness, we use the *mantle loading model with no density variation* (*f*=0). The sensitivity of the load ratio is investigated with the *mantle loading model with density variation*. For comparison, we also present results for the *density variation model without mantle loading* to distinguish which type of loading (mantle loading or density variation) is more critical in fitting observed admittance. These loading models are described in Text S1.

The purpose of the loading model is to infer lithospheric parameters of the basin area, such as elastic thickness, crustal density and load ratio. Densities of the crust and mantle were assumed in deriving the crustal thickness model (Wieczorek et al., 2022). Therefore, crustal density ($\rho_c$) and mantle density ($\rho_m$) are consistent with the W2022 crustal thickness model. For impact basins located in the southern highlands (Hellas, Argyre and Isidis), the crustal thickness ($T_c$)) in the loading model is 63 km and the crustal density ($\rho_c$) is the same as $\rho_{c,south}$. In the Utopia basin, the crustal thickness ($T_c$) is 37 km and the crustal density ($\rho_c$) is fixed at 2900 kg/m³. Other loading parameters are fixed and listed in Table 1.

**Table 1.** Fixed parameters in the loading model.

| Symbol | Description | Value | Unit |
|---|---|---|---|
| $R_{mp}$ | Mean planetary radius | 3396 | km |
| $M$ | Mass of Mars | $6.41712 \times 10^{23}$ | kg |
| $g_{sur}$ | Surface acceleration | 3.71 | m/s² |
| $\rho_c$ | Crust density | Utopia: 2900 | kg/m³ |
|  |  | Hellas, Argyre, and Isidis: equals to $\rho_{c,south}$ |  |
| $\rho_m$ | Mantle density | 3382 | kg/m³ |
| $T_c$ | Crust thickness | Utopia: 37 | km |
|  |  | Hellas, Argyre, and Isidis: 63 |  |
| $v$ | Poisson ratio | 0.25 |  |
| $E$ | Elastic module | 100 | GPa |

To study the sensitivity of elastic thickness, we set the load ratio to 0. This corresponds to *mantle loading without density variation* (Text S1). For the four W2022 crustal thickness models with $\rho_{c,south}$ ranging from 2600 to 2900 kg/m³, we calculated the modeled admittance with $T_e$ of 0, 20, and 100 km. The results for the Argyre basin are shown in Figure 4.

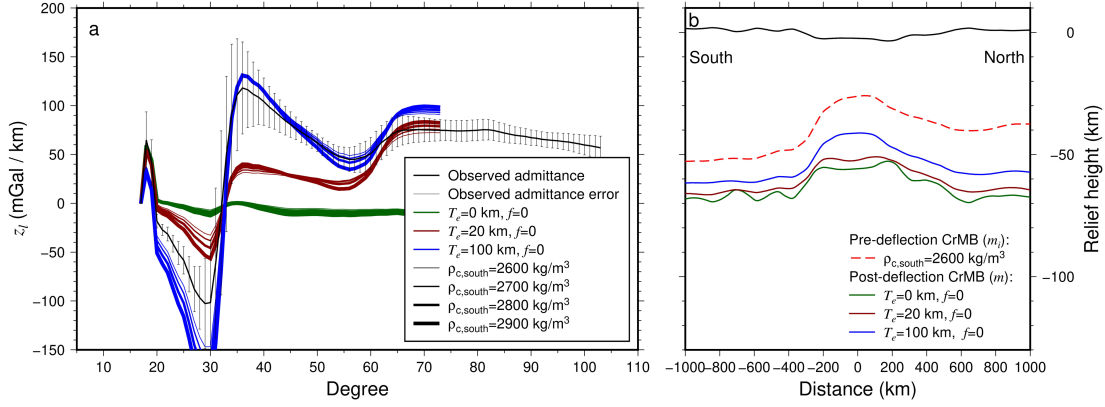

**Figure 4.** (a) Comparison of observed and modeled admittance of *mantle plug loading model with no density variation* for the Argyre basin. Different elastic thicknesses (0, 20, and 100 km) and crustal thickness models with different values of $\rho_{c,south}$ (2600, 2700, 2800, and 2900 kg/m³) are shown. The load ratio is set to 0. (b) Post-deflection sub-surface CrMB relief for different elastic thicknesses (0, 20, and 100 km, solid line). Pre-deflection CrMB relief ($m_i$) is from crustal thickness model with $\rho_{c,south}$ = 2600 kg/m³ (Red dashed line).

In Figure 4a, the observed admittance (black solid line) is shown with 3 times admittance error $\sigma_z(l)$. Since the crustal density ($\rho_c$) is same of $\rho_{c,south}$, lower $\rho_{c,south}$ also means that the crustal density is lower and thus crustal-mantle density contrast is greater. As can be seen from Figure 4a, the elastic thickness of lithosphere affects the modeled admittance. Models with larger elastic thicknesses (100 km, blue solid curves) are better fit to the observed admittance compared to smaller elastic thicknesses (0 and 20 km, green and red curves). At the same elastic thickness, the modeled admittance value is insensitive to the crustal thickness model we used.

To investigate the effect of elastic thickness variation on lithospheric deflection, we expanded pre- and post- CrMB relief ($m_i$ and $m$) from spherical harmonics into a spatial grid, and then captured profiles at the Argyre basin for comparison. Figure 4b shows the pre-CrMB relief ($m_i$) for the crustal thickness model with $\rho_{c,south}$=2600 kg/m³ (red dashed line) and the post-deflection CrMB relief ($m$) predicted from the loading model with $T_e$ of 0, 20, and 100 km (green, red, and blue solid lines).

In our mantle loading model, lithospheric deflection ($w$) is indicated by the difference between $m_i$ and $m$. Small elastic thickness (green solid line for 0 km $T_e$) corresponds to large lithosphere ($w$). Larger elastic thickness (blue solid line for 100 km $T_e$) results in less lithospheric deflection, so the difference between $m_i$ and $m$ is small.

In addition, the sensitivity analysis of load ratio ($f$) is performed using the *mantle loading model with density variation* (Text S1). In our study, we simplified the density variation loading by Belleguic et al. (2005) by setting the thickness of this density layer ($B$) to be the same as the crustal thickness ($T_c$). This means that lateral density variations are assumed to originate within the crust. If this density anomaly within the crust is positive, it implies that intrusive-like structures may be present in the region (Belleguic et al., 2005) or that there is infilling of high-density volcanic material on the surface (Ding et al., 2019). By contrast, the negative density anomaly indicates that the region may be affected by the impact-generated pore formation process (Milbury et al., 2015; Soderblom et al., 2015; Ding et al., 2018), resulting in higher porosity and lower crust density.

The comparison between observed and modeled admittance is shown in Figure 5. Each panel indicates the combination of the crustal thickness model with $\rho_{c,south}$ =2600 and 2900 kg/m$^3$ and $T_e$ of 0, 20, and 100 km. Modeled admittance curves in each panel have different load ratio values (0, 0.1, 1, and 2). We only calculated modeled admittance for the positive value of the load ratio as singular values may occur when it is negative (Beuthe et al., 2012).

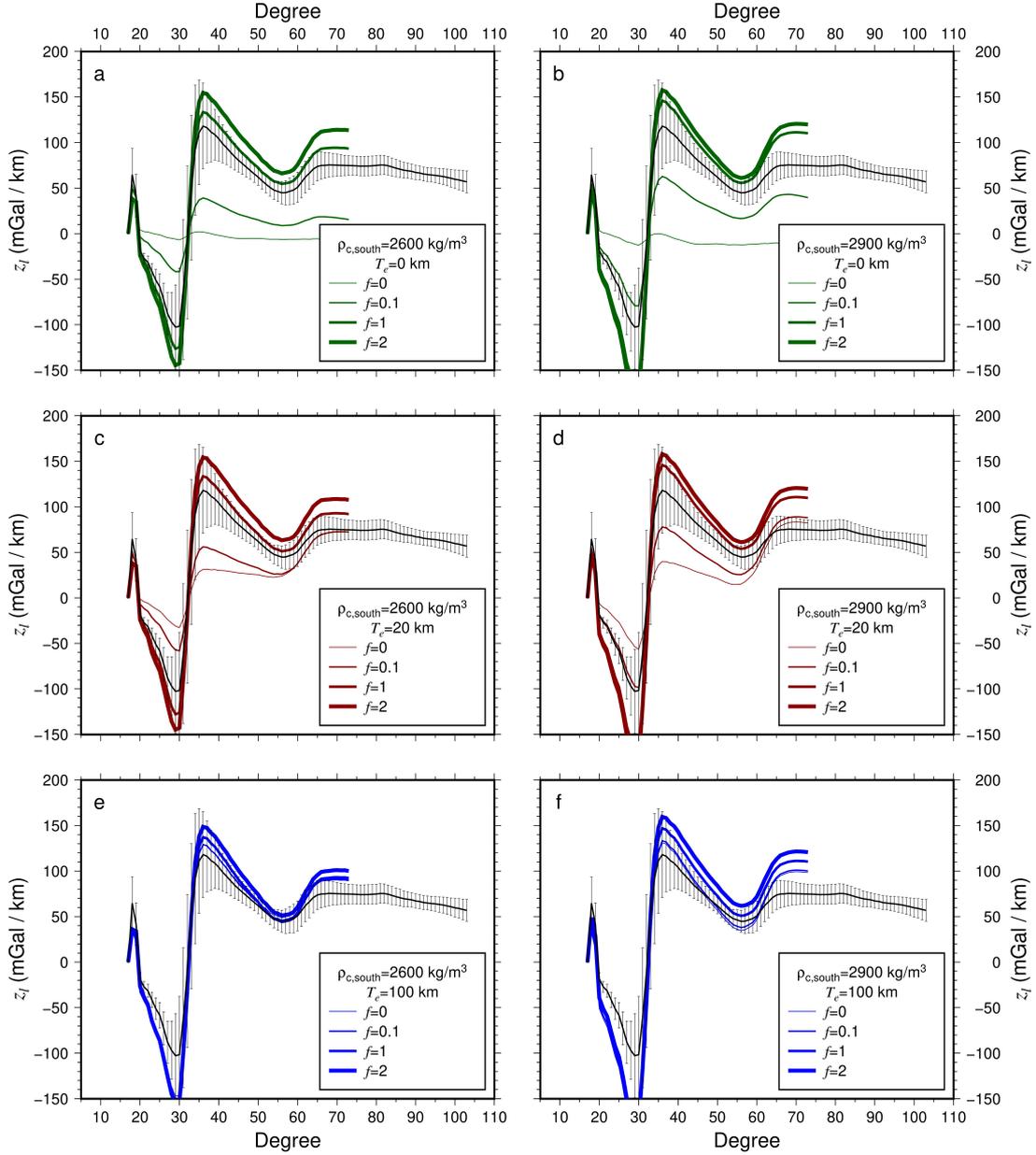

**Figure 5.** Comparison between observed admittance and modeled admittance of *mantle plug loading model with density variation* for different value of load ratio values ($f$=0, 0.1, 1, and 2). (a) $T_e$ = 0 km and $\rho_{c,south}$=2600 kg/m³. (b) $T_e$ = 0 km and $\rho_{c,south}$=2900 kg/m³. (c) $T_e$ = 20 km and $\rho_{c,south}$=2600 kg/m³. (d) $T_e$ = 20 km and $\rho_{c,south}$=2900 kg/m³. (e) $T_e$ = 100 km and $\rho_{c,south}$=2600 kg/m³. (f) $T_e$ = 100 km and $\rho_{c,south}$=2900 kg/m³.

For 0 km and 20 km elastic thicknesses (Figure 5a, 5b, 5c, and 5d), model of $f$=0 do not fit the observed admittance. Nevertheless, as the load ratio increases, modeled admittance gradually approaches the observed admittance. We find that when considering lateral density variation, model with small elastic thickness can achieve good fit compared to those without it.

To distinguish which type of loading (mantle loading or density variation) is most critical in the fitting of observed admittance, sensitivity analysis of the *density variation without mantle loading model* (Text S1) was performed (Figure 6). In this case, we calculated the admittance value with crustal density ($\rho_c$) of 2600 and 2900 kg/m$^3$. The density variation loading model does not reconstruct the mascon feature that admittance shifts sharply from negative to positive in a narrow spherical harmonic band. The fit of the modeled admittance to the observed admittance is relatively poor compared to the *mantle plug loading model with no density variation* (Figure 4). This suggests that modelling the mantle plug structure is the key to the fit of observed admittance in the Argyre region.

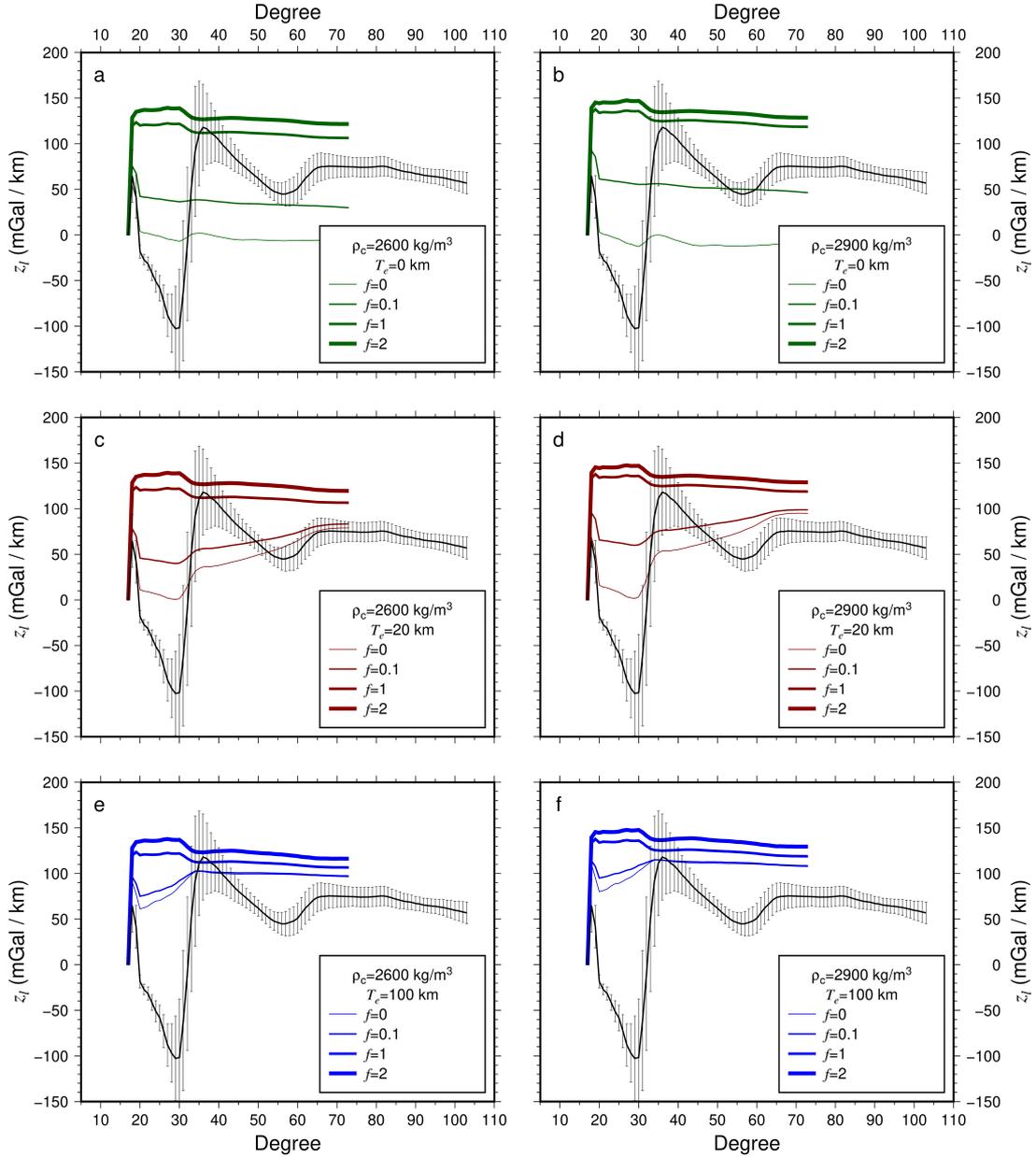

**Figure 6.** Comparison between observed admittance and modeled admittance of *density variation loading model* without mantle plug for different load ratio values ($f$=0, 0.1, 1, and 2). (a) $T_e$=0 km and $\rho_c$=2600 kg/m$^3$. (b) $T_e$=0 km and $\rho_c$=2900 kg/m$^3$. (c) $T_e$=20 km and $\rho_c$=2600 kg/m$^3$. (d) $T_e$=20 km and $\rho_c$=2900 kg/m$^3$. (e) $T_e$=100 km and $\rho_c$=2600 kg/m$^3$. (f) $T_e$=100 km and $\rho_c$=2900 kg/m$^3$.

From the results of parameter sensitivity analysis, we find that the effect of load ratio on the modeled admittance becomes weak if the elastic thickness of the lithosphere is large (Figure 5e and 5f). At the same time, modeled admittances with large value load ratios are similar (thicker lines in

Figure 5), indicating that variations in elastic thickness on the modeled admittances also become insensitive under higher load ratio conditions.

For the former, the obvious reason is that the flexural rigidity (Equation 2) of the lithosphere is proportional to the elastic thickness. Under the condition of large elastic thickness, the greater flexural rigidity of the lithosphere can resist deflection. Therefore, the modeled admittance is insensitive to changes in load ratio (*f*) at larger elastic thickness conditions.

For the matter that modeled admittance is insensitive to changes in elastic thickness under relatively large load ratios, we further examine the relationship between load ratio (*f*) and lithospheric deflection (*w*). We expand post-deflection CrMB (*m*) obtained under different load ratio conditions from spherical harmonics to spatial grids and extract profiles for comparison. In Figure 7, we present the post-deflection CrMB relief (*m*) for different values of load ratio (*f*) with 20 km $T_e$, as well as the pre-deflection CrMB relief (mi) from crustal thickness model of $\rho_{c,south}$=2600 kg/m³. Lithospheric deflection (w) is defined as the difference between m and mi.

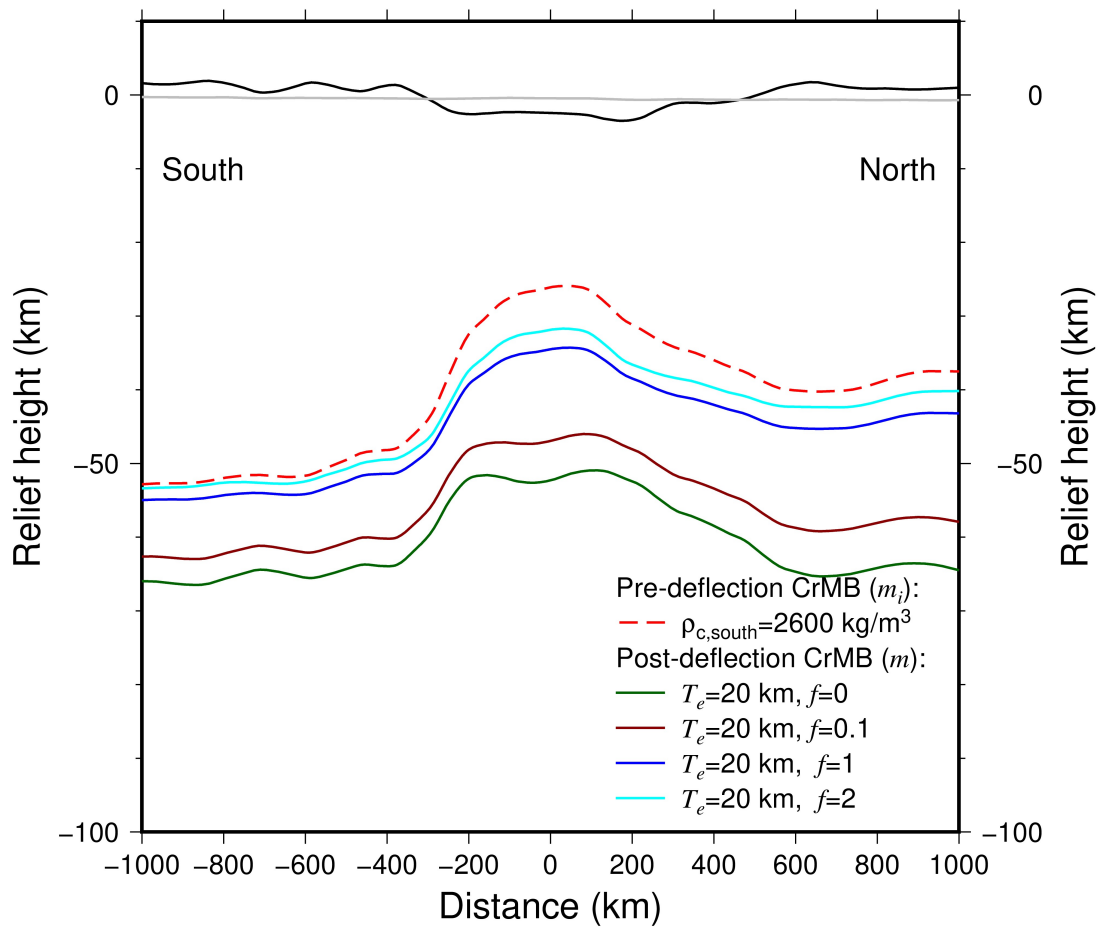

**Figure 7.** Post-deflection sub-surface CrMB relief for different load ratio values (*f*). Elastic thickness is assumed to be 20 km and crustal thickness model is with $\rho_{c,south}$= 2600 kg/m$^3$.

In Figure 7, we find that post-deflection CrMB relief (m, green solid line in Figure 7) is below pre-deflection CrMB relief ($m_i$, red dashed line in Figure 7) for load ratio *f*=0. This suggests that the net load of topography and mantle uplift in the Arygre basin region is positive (causing downward deflection). In other words, the mass surplus of mantle plug excess negative loads of depressed basal topography. As the load ratio (*f*) increases, the magnitude of lithospheric deflection (*w*) decreases. This indicates that the increase in the load ratio reduces the net load on the lithosphere.

We expand density variation (Δ$\rho$) at different load ratio values to spatial grids (Figure 8). We find that in the Argyre region, the positive load ratio (*f* > 0) corresponds to a negative density anomaly. Therefore, as the load ratio increases, the effect of negative density variation gradually increases, reducing both the net load and the resulting deflection (*w*). This explains the insensitivity of the model admittance to changes in elastic thickness under large load ratios. As the load ratio increases, initial positive load from topography and mantle plug are counteracted by lateral negative density anomaly. Under this condition of large load ratio, the magnitude of the net load is small, so the deflection of the lithosphere is subtle whether $T_e$ is large or small. Therefore, the sensitivity of the model admittance to changes in elastic thickness is weak.

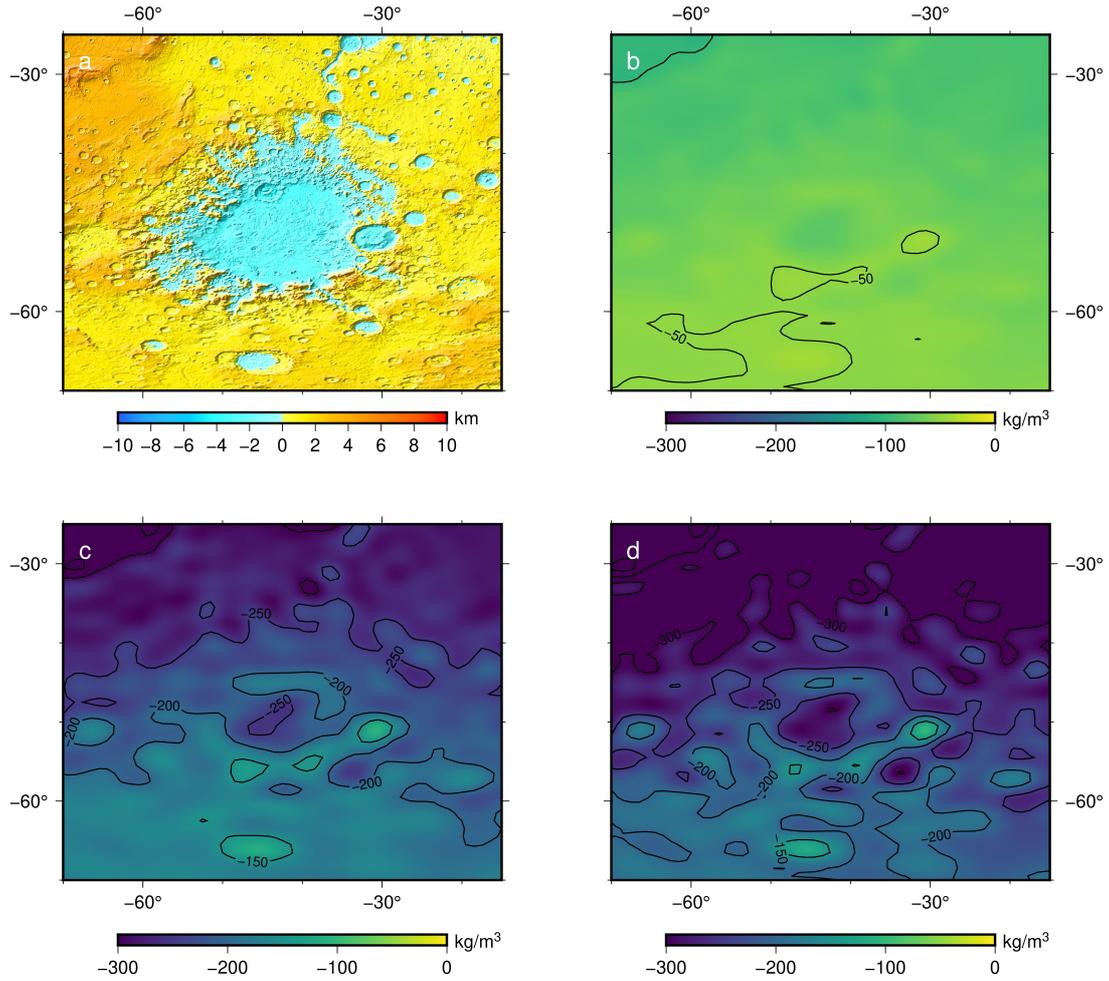

**Figure 8.** (a) Topography of the Arygre basin region, reference to geoid. (b-d) Density variation ($\Delta\rho$) expanded in spatial domain for $f$=0.1, 1, and 2. Elastic thickness $T_e$=20 km and $\rho_{c,south}$= 2600 kg/m$^3$ in crustal thickness model. Lateral negative density variation increases with load ratio $f$.

Through the above sensitivity analysis of elastic thickness and load ratio at the Arygre basin, we found that elastic thickness significantly affects the magnitude of lithospheric deflection and model admittance for mantle loading model if density variation is not considered (Figure 4) or load ratio is relatively small (thinner lines in Figure 5). Under large load ratio, the modeled admittance is insensitive to $T_e$ change (thicker lines in Figure 5). For large $T_e$ values (Figure 5e and 5f), the dependence of the modeled admittance on load ratio change is also weak. Although the density variation loading model without the mantle plug can also approximate the observed admittance to some extent (Figure 6), modelling the load of the mantle plug is the key to obtaining a better fit (Figures 4 and 5).

Considering density variation, the load ratio substantially affects lithospheric deflection

(Figure 7). Comparison of CrMB relief before and after deflection indicates that net load in the basin area is positive when load ratio $f = 0$, and this net load decreases with increasing load ratio (Figure 7). This is because the positive load ratio corresponds to the negative density anomaly (Figure 8), which reduces the positive net load. Sensitive analysis results for Isidis, Hellas, and Utopia are similar.

3.3 Elastic thickness and load ratio for Arygre and Isidis basin

To quantify the range of elastic thickness and load ratios that can be constrained by observed gravity and topography data, we calculated the misfit between observed and modeled admittance from a priori range of $T_e$ and $f$. Elastic thickness is sampled in 1 km steps within 0-350 km. The load ratio is sampled with an interval of 0.01 from 0 to 2. The normalized misfit ($\sigma$) for each parameter set is expressed with chi-squared function with 3 admittance errors (Ding et al., 2019; Zhong et al., 2022):

$$\sigma^2 = \frac{1}{l_{max}-l_{win}+1}\sum_{l=l_{win}}^{l_{max}}\left[\frac{z^{obs}(l)-z^{mod}(l)}{3\sigma_z(l)}\right]^2 \quad (16)$$

in which $l_{max}$ represents the maximum fit degree, which is calculated by the degree strength of Mars gravity field ($l_{strength}$ =90) minus the localized bandwidth: $l_{max} = l_{strength} - l_{win}$. $z^{obs}(l)$ is the admittance function calculated by the gravity field model and topography data and $z^{mod}(l)$ is the admittance function from the loading model. The expected value of the chi-square distribution is 1, so we set the threshold of misfit to 1.

The distribution of misfit with respect to elastic thickness and load ratio in the priori range for the Argyre basin is shown in Figure 9. Results from different crustal thickness models are shown in Figure 9a to 9d. For the Arygre basin, the crustal thickness model with $\rho_{c,south}$=2600 kg/m³ yields the largest parameter area with misfit <1. The distribution of model parameters with misfit <1 gives an elastic thickness of 47.3±23.8 km and a load ratio of 0.59±0.31 (Figure 9e and 9f). The model with the smallest misfit is located near $T_e$~0 km and $f$~0.6. For comparison, we calculated the modeled admittance and correlation for the two models with $T_e$= 0 km and $f$=0.6 versus $T_e$=55 km and $f$=0 (Figure 9g and Figure 9h). The former is the smallest misfit model with the best load ratio for 0 km $T_e$, while the latter is the best model for mantle loading without density variation ($f$=0). These two models are two end member cases for the mantle loading model. We found that the model admittance and correlation from both models fit the observed data well.

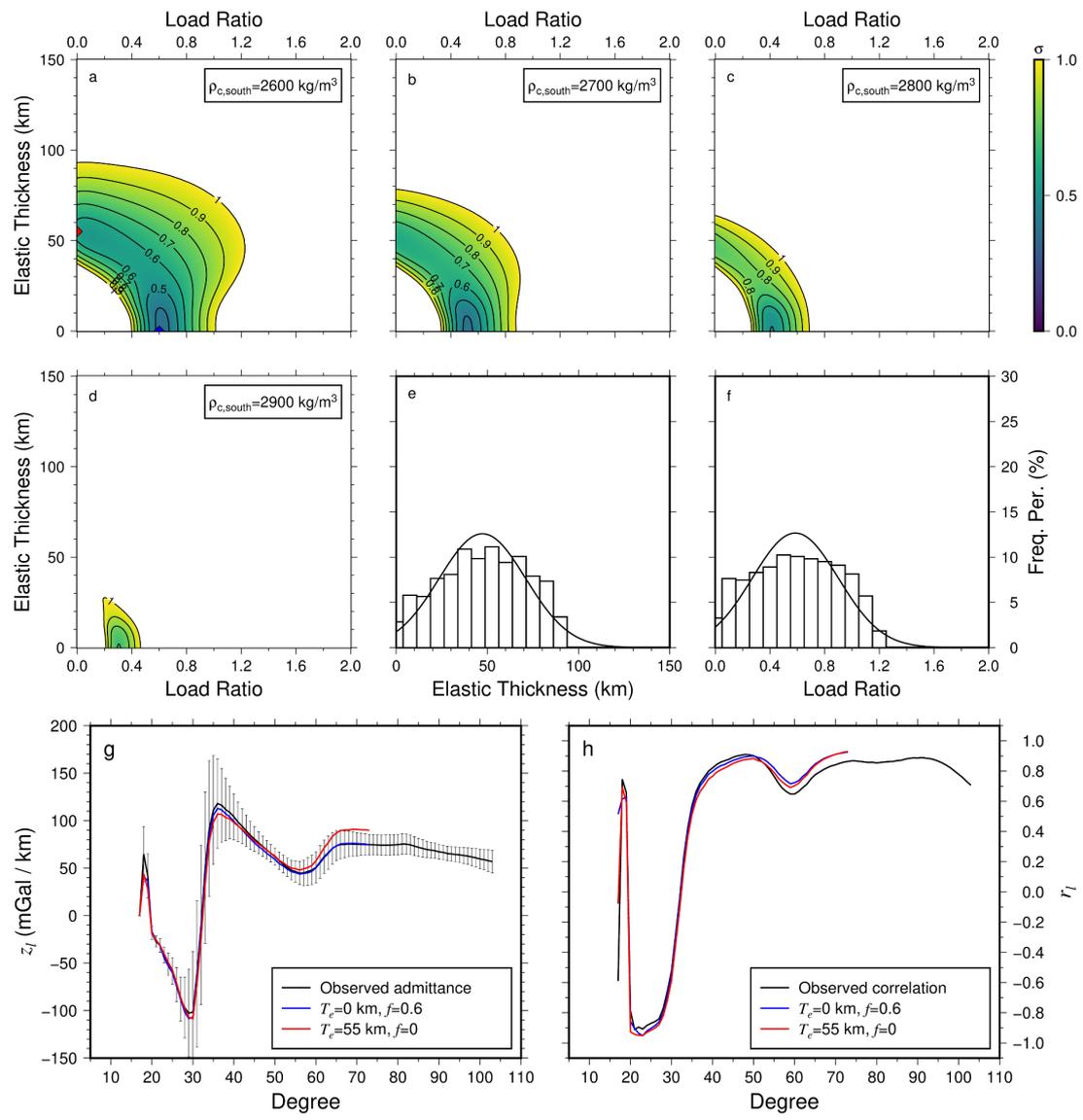

**Figure 9.** (a-d) Distribution of misfit with respect to elastic thickness and load ratio for the Argyre basin with different crustal thickness models ($\rho_{c,south}$ from 2600 to 2900 kg/m$^3$). (e) Histogram of elastic thickness constrained by misfit <1. The elastic thickness is estimated as 47.3±23.8 km. (f) Histogram of load ratio constrained by misfit <1. The load ratio is estimated as 0.59±0.31 km. (g) Comparison between observed and modeled admittance of two end-member parameter combinations: $T_e$=0 km and $f$=0.6 versus $T_e$=55 km and $f$=0. (h) Comparison between observed correlation and modeled correlation of two end-member parameter combinations.

The misfit distribution for the Isidis basin (Figure 10) indicates that the crustal thickness model with $\rho_{c,south}$=2900 kg/m$^3$ contains the most models with misfit < 1. The elastic thickness is estimated as 74.2±45.5 km. Models with misfit < 1 involve a wide range of load ratios (Figure 10d) and parameter distributions show that modeled fit is not sensitive to load ratio variation (Figure 10f). The load ratio is estimated as 1.16±0.55. The best model for 0 km elastic thickness has a load ratio of ~0.55, while the best model for mantle loading without density variation has an elastic thickness of ~70 km. Their modeled admittance and correlation are shown in Figure 10g and Figure 10h. For the Isidis basin, the 0 km $T_e$ model with $f$=0.55 has better fitting results to observed data than the 70 km $Te$ model with zero load ratio. This suggests that lateral density variation needs to be considered in the Isidis basin region.

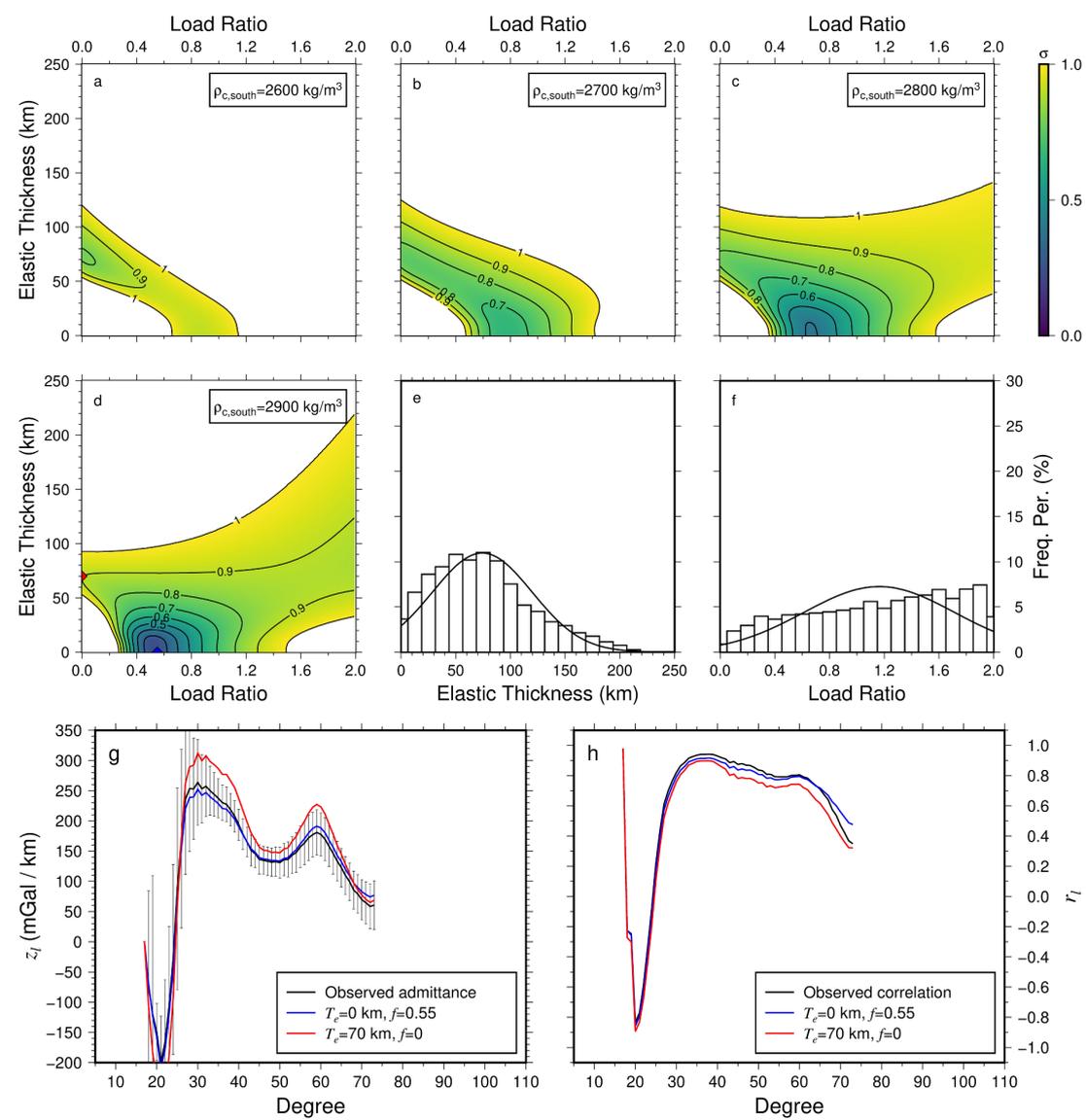

**Figure 10.** (a-d) Distribution of misfit with respect to elastic thickness and load ratio for the Isidis basin with different crustal thickness models ($\rho_{c,south}$ from 2600 to 2900 kg/m³). (e) Histogram of elastic thickness constrained by misfit <1. The elastic thickness is estimated as 74.2±45.5 km. (f)

Histogram of load ratio constrained by misfit <1. The load ratio is estimated as 1.16±0.55. (g) Comparison between observed and modeled admittance of two end-member parameter combinations: $T_e$=0 km and $f$=0.55 versus $T_e$=70 km and $f$=0. (h) Comparison between observed correlation and modeled correlation of two end-member parameter combination.

## 4. Discussion

4.1 Spectral analysis of admittance for impact basin

The loading state in the impact basin region is complex. Mantle uplifted structure significantly influences local gravity signals, and also exerts loading on the lithosphere. In the impact basin region, lithospheric model that do not consider mantle loading are difficult to reconcile with the actual loading state, even though loading from topography and density variation have been modeled. In Ding et al. (2019), the spatial domain method was used to estimate $T_e$ at the impact basin because the modeled admittance obtained from the loading model differed substantially from the observed admittance in the spectral domain. In this study, with the help of mantle loading model integrated with the latest crustal thickness model, we obtained a good fit for the admittance and correlation spectrum of impact basins (Figure 9g and 9h, Figure 10g and 10h). This demonstrates the applicability of mantle loading on Mars impact basin. Lithospheric parameters in loading model are well constrained in the Argyre and Isidis basin. In the Hellas and Utopia basin, the negative to positive trend in the observed admittance is not significant, so that the elastic thickness and load ratio are not bounded (Figures S1 and S2).

4.2 Trade-off between elastic thickness and load ratio

In sensitivity analysis and misfit distribution, our mantle loading model with density variation exhibits a trade-off between elastic thickness and load ratio. In the Arygre and Isidis basins, the mantle loading model without density variation ($f$=0, large $T_e$) is comparable to the model with 0 km $T_e$ and large $f$. Our model calculation results show that considering lateral density variation in the mantle loading model can improve model fit, but do not indicate whether actual loading of the impact basin is significantly influenced by lateral density variation. To better constrain the elastic thickness of the impact basin, it is necessary to place external constraints on the load ratio.

Impact-generated pore formation may be the key to modeling lateral density variation in the impact basin region. In this study, positive load ratio corresponds to negative density anomaly, consistent with impact-induced porosity amplification. Porosity changes around large impact basins have been suggested in studies for large impact basins on the Moon (Soderblom et al., 2015; Wahl et al., 2020; Huang et al., 2022). The vertical porosity structure (Besserer et al., 2014) is also integrated into crustal thickness model of Mercury (Beuthe et al., 2020). Therefore, porosity structure in the crust of Mars can provide information on lateral density variation in the impact basin region, so that the load ratio in the mantle loading model can be better constrained, thus further limiting the range of estimated elastic thickness.

4.3 Elastic thickness in the impact basin region

In this study, elastic thickness estimates for the Argyre and Isidis basins are 47.3 km and 74.2 km, respectively. These results are within the range of the elastic thickness of the lithosphere from the thermal structure modeling results of Searls et al. (2006), which give a lithospheric thickness of ~34 km and ~70 km at 4 Ga. In previous studies on $T_e$ in the Isidis basin, Ritzer and Hauck (2009) gave results of 100-180 km, and Ding et al. (2019) suggest $T_e$ at Isidis >100 km. Isostatic compensation of topography is considered in their work, so similar results are obtained by them. Mancinelli et al. (2015) inferred $T_e$ ~36 km in the Isidis basin based on crustal structure obtained from impact simulation. Its modelling of mantle structure does not consider the key step in the mascon formation process: the uplift of the basin floor due to the difference between the compensation states of the internal and external region (Melosh et al., 2013; Freed et al., 2014). Our model uses the latest crustal thickness model of Mars to minimize the uncertainty caused by crustal density and mantle density, which is the advantage of our model. Nevertheless, the mantle loading model proposed in this paper also has limitations. That is, it does not model the loading of in-filling material.

In previous studies of impact basins using loading models without considering an initial mantle plug, the estimated $T_e$ value was relatively small (McGovern et al., 2002, Wagner et al., 2022). Small $T_e$ in impact basin regions is usually interpreted as a temporary reduction in lithosphere strength caused by impact heating. Nevertheless, one of the conditions for mascon basin formation is rapid cooling of the lithosphere (Melosh et al., 2013; Freed et al., 2014), which means that the lithosphere must already have a certain thickness during post-impact adjustment when the load is acting. Therefore, temporary lithosphere heating induced by the impact process over a long geologic period should be relatively weak in a mascon basin. This suggests that the larger elastic thickness obtained by the loading model considering CrMB structures at the impact basin (Riter and Hauck et al., 2009; Ding et al., 2019; and this work) is more consistent with the actual formation scenario of the mascon basin.

## 5. Conclusion

In this work, the proposed lithospheric mantle loading model with lateral density variation uses the crustal thickness model to construct the CrMB relief, which reduces the uncertainty involved in the isostatic hypothesis and crustal density. Our newly proposed loading model achieved a good fit for observed admittance. It also suggests that modelling mantle loading is more critical than modelling density variation in the impact basin region. This mantle loading model is applicable to regions that exhibit typical mascon admittance characteristics (with low gravity-topography correlation and sharp admittance transition).

Calculation in the Argyre and Isidis basins shows that the mantle loading model can fit the observed admittance well. Elastic thickness of the Argyre and Isidis basins is estimated to be 47.3 and 74.2 km, consistent with the lithospheric elastic thickness derived from thermal modelling of the Martian lithosphere. Due to the trade-off between Te and f, no further constrains can be placed on the lateral density of the basin region.

**Acknowledgment**

We are grateful to M. Wieczorek for providing the open-source software SHTools (Wieczorek et al., 2018). F. Li is supported by the National Natural Science Foundation of China (42030110, 41874010). J. G. Yan is supported by the National Key R&D Program (2022YFF0503200, 2022YFF0503202), National Natural Science Foundation of China (U1831132), the Fundamental Research Funds for the Central Universities (2042019kf0191) and Innovation Group of Natural Fund of Hubei Province (2018CFA087). Q. Y. Deng is supported by the Fundamental Research Funds for the Central Universities (2042022kf1004). J.-P. Barriot is funded by a DAR grant in planetology from the French Space Agency (CNES). M. Ye is supported by the National Natural Science Foundation of China (41804025). Several figures were created with the Generic Mapping Tools (GMT) software (Wessel & Smith, 1991). Q. Y. Deng thanks Z. Y. Luo for her passionate love.

**Open Research**

Data Availability Statement

The MRO gravity field models of Mars are available at Zuber et al. (2010). The crustal thickness model of Mars is from Wieczorek (2022). The 719-degree and 2600-degree spherical harmonic shape models of Mars can be found in Wieczorek (2007).

# Supporting Information for
# Lithospheric loading model for large impact basin where mantle plug presents

Qingyun Deng[1], Zhen Zhong[2], Mao Ye[1*], Wensong Zhang[1], Denggao Qiu[1], Chong Zheng[1], Jianguo Yan[1], Fei Li[1,3], Jean-Pierre Barriot[1,4]

[1]State Key Laboratory of Information Engineering in Surveying, Mapping and Remote Sensing, Wuhan University, Wuhan, China,

[2]School of Physics and Electronic Science, Guizhou Normal University, Guiyang, China,

[3]Chinese Antarctic Center of Surveying and Mapping, Wuhan University, Wuhan, China,

[4]Geodesy Observatory of Tahiti, Tahiti, French Polynesia

*Corresponding author: Mao Ye (mye@whu.edu.cn)

**1. Topography loading, mantle loading, and density variation loading**

We started from the relation between lithospheric deflection ($w$) and net vertical pressure ($q$):

$$w = \alpha q \quad (S1)$$

where $\alpha$ is the compensation ratio:

$$\alpha = \frac{n(n+1) - 1 + v}{\sigma[n^3(n+1)^3 - 4n^2(n+1)^2] + \tau[n(n+1) - 2]} \quad (S2)$$

and

$$\sigma = \frac{D}{R^4} = \frac{ET_e^3}{12R^4(1-v^2)} \quad (S3)$$

$$\tau = \frac{ET_e}{R^2} \quad (S4)$$

The net vertical pressure in **topography loading** model is written as:

$$q = g_1 \rho_c h - g_2(\rho_m - \rho_c)w \quad (S5)$$

Then the relation between the deflection ($w$) and topography ($h$) is given as:

$$w = \frac{\alpha g_1 \rho_c}{1 + \alpha g_2(\rho_m - \rho_c)} h \quad (S6)$$

If $T_e$=0 km, then the relationship is reduced to:

$$w = \frac{g_1 \rho_c}{g_2(\rho_m - \rho_c)} h \quad (S7)$$

In the **density variation loading model**, the net vertical pressure is expressed as:

$$q = g_1 \rho_c h - g_2(\rho_m - \rho_c)w + g_3 \Delta\rho B \quad (S8)$$

where the relationship between the initial topography loading ($h_i$) and the density variation layer is indicated by the load ratio $f$:

$$f = -\frac{\Delta\rho B}{\rho_c h_i} \quad (S9)$$

The initial topography loading ($h_i$) is the pre-deflection topography relief, and the present-day topography ($h$) takes into account the lithospheric deflection ($w$):

$$h = h_i - w \tag{S10}$$

Equation S1 can be re-written as:

$$w = \alpha g_1 \rho_c h - \alpha g_2 (\rho_m - \rho_c) w + \alpha g_3 \Delta\rho B \tag{S11}$$

Re-arranging Equation S11 and Substituting $w$ with ($h_i$ - $h$):

$$[1 + \alpha g_2 (\rho_m - \rho_c)](h_i - h) = \alpha g_1 \rho_c h + \alpha g_3 \Delta\rho B \tag{S12}$$

Re-arranging Equation S12 and substituting hi with $h_i = -\Delta\rho B / \rho_c f$ :

$$\frac{-\Delta\rho B[1 + \alpha g_2(\rho_m - \rho_c)]}{\rho_c f} = [1 + \alpha g_2(\rho_m - \rho_c) + \alpha g_1 \rho_c]h + \alpha g_3 \Delta\rho B \tag{S13}$$

Then the density variation can be expressed as:

$$\Delta\rho = \frac{-\rho_c f}{B} \frac{1 + \alpha g_2(\rho_m - \rho_c) + \alpha g_1 \rho_c}{1 + \alpha g_2(\rho_m - \rho_c) + \alpha f g_3 \rho_c} h \tag{S14}$$

Re-arranging Equation S11 and substituting $\Delta\rho B$ with $-f\rho_c h_i$:

$$w = \alpha g_1 \rho_c h - \alpha g_2 (\rho_m - \rho_c) w - \alpha f g_3 \rho_c h_i \tag{S15}$$

Substituting $h_i$ with $h + w$, we have the relationship of:

$$w = \alpha g_1 \rho_c h - \alpha g_2 (\rho_m - \rho_c) w - \alpha f g_3 \rho_c (h + w) \tag{S16}$$

Re-arranging this Equation, we have:

$$w = \frac{\alpha g_1 \rho_c - \alpha f g_3 \rho_c}{1 + \alpha g_2(\rho_m - \rho_c) + \alpha f g_3 \rho_c} h \tag{S17}$$

If $f$=0, then Equation S17 reduced to Equation S6. If $T_e$=0 km, then Equation S14 and S17 reduced to:

$$\Delta\rho = -\frac{\rho_c f}{B} \frac{g_2(\rho_m - \rho_c) + g_1 \rho_c}{g_2(\rho_m - \rho_c) + f g_3 \rho_c} h \tag{S18}$$

$$w = \frac{g_1 \rho_c - f g_3 \rho_c}{g_2(\rho_m - \rho_c) + f g_3 \rho_c} h \tag{S19}$$

In the **mantle loading model without density variation**, we modeled the post-impact mantle plug as:

$$q = g_1 \rho_c h + g_2 (\rho_m - \rho_c) m \tag{S20}$$

where *m* is the post-deflection CrMB relief. The relation between lithospheric deflection (*w*), pre-, and post-deflection CrMB ($m_i$ and *m*) is:

$$m = m_i - w \tag{S21}$$

where $m_i$ is the pre-deflection CrMB relief after impact.

Substituting Equation S21 to Equation S20, we have:

$$m_i - m = \alpha g_1 \rho_c h + \alpha g_2 (\rho_m - \rho_c) m \tag{S22}$$

We can get the relationship between *m*, $m_i$, and *h*:

$$m = \frac{-\alpha g_1 \rho_c}{1+\alpha g_2(\rho_m-\rho_c)} h + \frac{1}{1+\alpha g_2(\rho_m-\rho_c)} m_i \tag{S23}$$

If $T_e$=0 km, then we have 1/$\alpha$=0. Equation S23 is expressed as:

$$m = -\frac{g_1 \rho_c}{g_2(\rho_m-\rho_c)} h \tag{S24}$$

This express of post-deflection CrMB is identical to lithosphere deflection (*w*) in topography loading model (*m* is measured positive upward, but *w* is measured positive downward).

If both of the density variation layer and the post-impact mantle plug are incorporated in calculation of net vertical pressure (**mantle loading model with density variation**), then we have:

$$w = \alpha g_1 \rho_c h + \alpha g_2 (\rho_m - \rho_c) m + \alpha g_3 \Delta\rho B \tag{S25}$$

From this Equation, we can derive the relationship of m and $\Delta\rho$ as linearly combination of $m_i$ and *h*. Substituting S21 to S25, we can eliminate the present-day CrMB relief *m*:

$$w = \alpha g_1 \rho_c h + \alpha g_2 (\rho_m - \rho_c)(m_i - w) + \alpha g_3 \Delta\rho B \tag{S26}$$

Re-arrange Equation S26 and substituting Equation S10:

$$[1 + \alpha g_2 (\rho_m - \rho_c)](h_i - h) = \alpha g_1 \rho_c h + \alpha g_2 (\rho_m - \rho_c) m_i + \alpha g_3 \Delta\rho B \tag{S27}$$

Re-arrange Equation S27 and substituting Equation S9:

$$-\frac{\Delta\rho B}{\rho_{cf}}[1 + \alpha g_2(\rho_m - \rho_c)] = [1 + \alpha g_2(\rho_m - \rho_c) + \alpha g_1 \rho_c] h + \alpha g_2(\rho_m - \rho_c) m_i +$$

$$\alpha g_3 \Delta\rho B \tag{S28}$$

Re-arranging Equation S28, we can solve the density variation with $h$ and $m_i$:

$$\Delta\rho = -\frac{\rho_c f}{B}\left(\frac{1+\alpha g_2(\rho_m-\rho_c)+\alpha g_1\rho_c}{1+\alpha g_2(\rho_m-\rho_c)+f\alpha g_3\rho_c}h + \frac{\alpha g_2(\rho_m-\rho_c)}{1+\alpha g_2(\rho_m-\rho_c)+f\alpha g_3\rho_c}m_i\right) \quad (S29)$$

To solve the post-deflection CrMB relief ($m$) with $h$ and $m_i$, we first eliminate the density variation term through substituting S9 to S25:

$$w = \alpha g_1\rho_c h + \alpha g_2(\rho_m-\rho_c)m - f\alpha g_3\rho_c h_i \quad (S30)$$

Then replace the initial topography loading $h_i$ with $h + w$:

$$w = \alpha g_1\rho_c h + \alpha g_2(\rho_m-\rho_c)m - f\alpha g_3\rho_c(h+w) \quad (S31)$$

Re-arranging Equation S31 and replacing $w$ with $m_i - m$:

$$[1+f\alpha g_3\rho_c](m-m_i) = [f\alpha g_3\rho_c - \alpha g_1\rho_c]h - \alpha g_2(\rho_m-\rho_c)m \quad (S32)$$

Then we have:

$$m = \frac{(f\alpha g_3\rho_c - \alpha g_1\rho_c)}{1+\alpha g_2(\rho_m-\rho_c)+f\alpha g_3\rho_c}h + \frac{(1+f\alpha g_3\rho_c)}{1+\alpha g_2(\rho_m-\rho_c)+f\alpha g_3\rho_c}m_i \quad (S33)$$

If $T_e$=0 km, Equation S29 and S33 reduce to:

$$\Delta\rho = -\frac{\rho_c f}{B}\left(\frac{g_2(\rho_m-\rho_c)+g_1\rho_c}{g_2(\rho_m-\rho_c)+fg_3\rho_c}h + \frac{g_2(\rho_m-\rho_c)}{g_2(\rho_m-\rho_c)+fg_3\rho_c}m_i\right) \quad (S34)$$

$$m = \frac{fg_3\rho_c - g_1\rho_c}{g_2(\rho_m-\rho_c)+fg_3\rho_c}h + \frac{fg_3\rho_c}{g_2(\rho_m-\rho_c)+fg_3\rho_c}m_i \quad (S35)$$

If $f$=0, then it means that the density variation loading is not considered. Equation S33 and S35 reduce to Equation S23 and S24.

**2. Misfit distribution for the Hellas and Utopia basin**

The observed admittance of the Hellas and Utopia basin does not show a pattern of sharp change from large negative to large positive within a narrow SH range. We speculate that this is why the mantle loading model is not applicable in both basins. In the misfit distribution of the Hellas basin, there are regions with small misfit in elastic thickness <50 km and load ratio between 0.6 and 1.6. Nevertheless, models with misfit <1 constitute a wide range for $T_e$ and $f$. This suggests that model performance is insensitive to elastic thickness. The calculation result for the Utopia basin is shown in Figure S2. Although the misfit value for the Utopia basin is relatively small, the sampling results do not provide a constraint on elastic thickness and *f*.

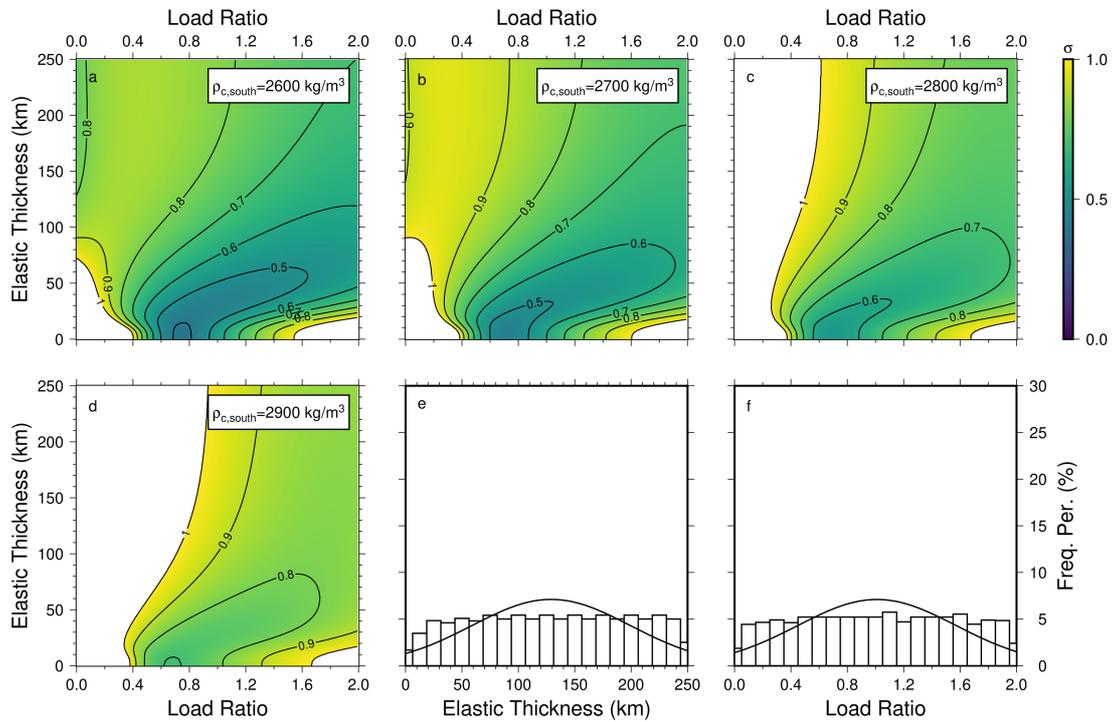

**Figure S1**. (a-d) Distribution of misfit with respect to elastic thickness and load ratio for the Hellas basin with different crustal thickness models ($\rho_{c,south}$ from 2600 to 2900 kg/m$^3$). (e) Histogram of elastic thickness constrained by misfit <1. (f) Histogram of load ratio constrained by misfit <1.

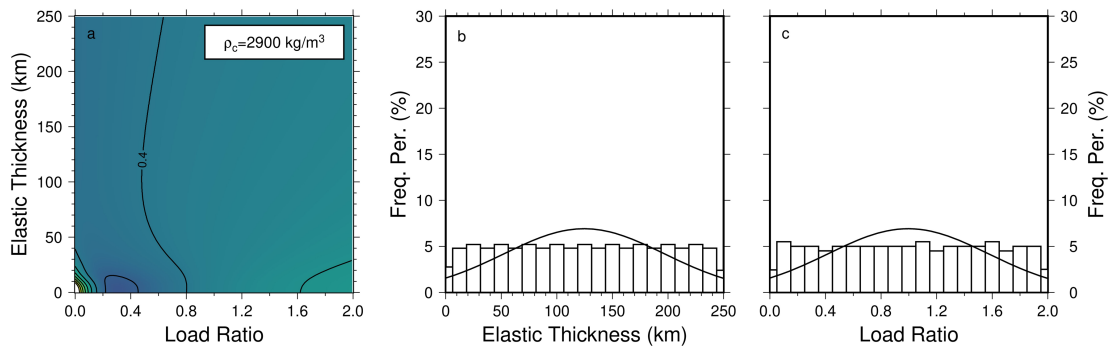

**Figure S2**. (a-d) Distribution of misfit with respect to elastic thickness and load ratio for the Utopia basin with different crustal thickness models ($\rho_{c,south}$ from 2600 to 2900 kg/m$^3$). (e) Histogram of elastic thickness constrained by misfit <1. (f) Histogram of load ratio constrained by misfit <1.